\documentclass[prl,twocolumn,superscriptaddress]{revtex4}%
\usepackage[ansinew]{inputenc}
\usepackage[dvips]{graphicx}
\usepackage{amsmath}
\usepackage{amsthm}
\usepackage{layout}
\usepackage{float}
\usepackage{subfigure}
\usepackage{amsfonts}
\usepackage{amssymb}%
\begin{document}

\title{Entanglement Distribution Revealed by Macroscopic Observations}

\begin{abstract}
What can we learn about entanglement between individual particles in
macroscopic samples by observing only the collective properties of the
ensembles? Using only a few experimentally feasible collective properties, we
establish an entanglement measure between two samples of spin-1/2 particles
(as representatives of two-dimensional quantum systems). This is a tight lower
bound for the average entanglement between all pairs of spins in general and
is equal to the average entanglement for a certain class of systems. We
compute the entanglement measures for explicit examples and show how to
generalize the method to more than two samples and multi-partite entanglement.

\end{abstract}
\date{\today}%

%TCIMACRO{\TeXButton{auth1}{\author{Johannes Kofler}}}%
%BeginExpansion
\author{Johannes Kofler}%
%EndExpansion
%

%TCIMACRO{\TeXButton{aff1}{\affiliation{Institut f\"{u}%
%r Experimentalphysik, Universit\"{a}%
%t Wien, Boltzmanngasse 5, 1090 Wien, Austria}
%}}%
%BeginExpansion
\affiliation{Institut f\"{u}r Experimentalphysik, Universit\"{a}%
t Wien, Boltzmanngasse 5, 1090 Wien, Austria}
%EndExpansion
%

%TCIMACRO{\TeXButton{aff2}{\affiliation
%{Institut f\"ur Quantenoptik und Quanteninformation, \"Osterreichische Akademie der Wissenschaften,\\ Boltzmanngasse 3, 1090 Wien, Austria}%
%}}%
%BeginExpansion
\affiliation
{Institut f\"ur Quantenoptik und Quanteninformation, \"Osterreichische Akademie der Wissenschaften,\\ Boltzmanngasse 3, 1090 Wien, Austria}%
%EndExpansion
%

%TCIMACRO{\TeXButton{auth2}{\author{{\v C}aslav Brukner}}}%
%BeginExpansion
\author{{\v C}aslav Brukner}%
%EndExpansion
%

%TCIMACRO{\TeXButton{aff1}{\affiliation{Institut f\"{u}%
%r Experimentalphysik, Universit\"{a}%
%t Wien, Boltzmanngasse 5, 1090 Wien, Austria}
%}}%
%BeginExpansion
\affiliation{Institut f\"{u}r Experimentalphysik, Universit\"{a}%
t Wien, Boltzmanngasse 5, 1090 Wien, Austria}
%EndExpansion
%

%TCIMACRO{\TeXButton{aff2}{\affiliation
%{Institut f\"ur Quantenoptik und Quanteninformation, \"Osterreichische Akademie der Wissenschaften,\\ Boltzmanngasse 3, 1090 Wien, Austria}%
%}}%
%BeginExpansion
\affiliation
{Institut f\"ur Quantenoptik und Quanteninformation, \"Osterreichische Akademie der Wissenschaften,\\ Boltzmanngasse 3, 1090 Wien, Austria}%
%EndExpansion
%

%TCIMACRO{\TeXButton{maketitle}{\maketitle}}%
%BeginExpansion
\maketitle
%EndExpansion

Observation of quantum entanglement between increasingly larger objects is one
of the most promising avenues of experimental quantum physics. Eventually, all
these developments might lead to a full understanding of the simultaneous
coexistence of a macroscopic classical world and an underlying quantum realm.
Macroscopic samples typically contain $N\sim10^{23}$ particles. Because the
system's Hilbert space grows exponentially with the number of constituent
particles, a complete microscopic picture of entanglement in large systems
seems to be in general intractable. The question arises: What can we learn
about entanglement between constituent particles of macroscopic samples, if
only limited experimentally accessible knowledge about the samples is available?

There is a strong motivation in addressing this question because of recent
experimental progress in creating and manipulating entangled states of
increasing complexity, such as spin-squeezed states of two atomic
ensembles~\cite{Polzik}. In such experiments one typically measures only
expectation values of \textit{collective operators} of two separated samples.
It is known that the two samples of spins can be characterized as either
entangled or separable by measuring collective spin operators~\cite{Sorensen}.
Furthermore, such measurements are shown to be sufficient to determine
entanglement measures of Gaussian states~\cite{Molmer} and of a pair of
particles that are extracted from a totally symmetric spin state (invariant
under exchange of particles)~\cite{WangMolmer}. It appears that collective
operators cannot be used to fully characterize entanglement in composite
systems without strong requirements on the symmetry of the state.

Here we present a general and practical method for entanglement detection
between two samples of spins. It solely employs \textit{collective spin
properties of the samples} and works irrespective of the number of spin
particles constituting the samples and with no assumption about the symmetry
or mixedness of the state. The method is based on an entanglement measure
which is a \textit{tight lower bound for average entanglement between all
pairs of spins }belonging to the two samples. This measure is \textit{equal}
to the average entanglement for a certain class of systems which need not be
totally symmetric. We generalize the method to obtain the entanglement measure
between $M$ separated spin samples based on collective measurements. The
results apply for any entanglement monotone that is a convex measure on the
set of density matrices (e.g., concurrence~\cite{Wootters},
negativity~\cite{Zycz1998}, three-way tangle~\cite{Coffman}).

\begin{figure}[t]
\begin{center}
\includegraphics{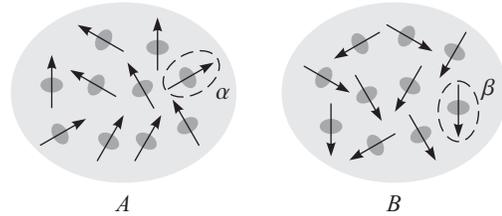}
\end{center}
\par
\vspace{-0.5cm}\caption{Two contiguous and non-overlapping spin subsystems $A$
and $B$, each of which contains a large number $n$ of spins. What can we learn
about entanglement of a pair $(\alpha,\beta)$ of spins chosen at random where
$\alpha\in A$ and $\beta\in B$, if individual spins are experimentally not
accessible but only the collective properties of the samples $A$ and $B$? What
can we learn about entanglement between $A$ and $B$ from such measurements?}%
\label{fig subsystems}%
\end{figure}

Consider two separated ensembles $A$ and $B$ of spin-$\frac{1}{2}$ particles
(figure~\ref{fig subsystems}). Each of them contains a large number $n$ of
spins. Because of the large dimensions ($d=2^{n}$) of the samples' Hilbert
spaces, the structure of entanglement between the two samples is considerably
more complex than between two single spins. While there are experimentally
viable methods for detecting entanglement, they still require a large number
of parameters to be determined (proportional to $d^{2}$~\cite{Horodecki}). The
problem simplifies in situations in which each of the ensembles of $n$ spins
can be treated as one large (total) spin of length $n/2$. This means that,
within the ensembles, the individual spin-$\frac{1}{2}$ particles form
symmetrized states (Dicke states). Though this reduces the dimension of the
Hilbert space of $A$ ($B$) to $d=n+1$, entanglement determination is still
demanding for large $n$ both experimentally and theoretically: analytical
solutions exist only for pure states in general and for mixed states only for
small $n$~\cite{Schliemann}.

In this paper, we give a method to detect entanglement between large spin
samples by measuring only a \textit{small number of collective spin
properties} (sample spin components and their correlations), which is
independent of the sample size $n$. The collective spin operators are%
\begin{equation}
\hat{S}_{i}^{A}\equiv\frac{\hbar}{2}\,%
%TCIMACRO{\dsum \nolimits_{\alpha\in A}}%
%BeginExpansion
{\displaystyle\sum\nolimits_{\alpha\in A}}
%EndExpansion
\,\hat{\sigma}_{i}^{(\alpha)},\;\;\hat{S}_{i}^{B}\equiv\frac{\hbar}{2}\,%
%TCIMACRO{\dsum \nolimits_{\beta\in B}}%
%BeginExpansion
{\displaystyle\sum\nolimits_{\beta\in B}}
%EndExpansion
\,\hat{\sigma}_{i}^{(\beta)}. \label{S}%
\end{equation}
The index $i$ denotes the spatial component of the spins: $i\in\{1\!\equiv
\!x,2\!\equiv\!y,3\!\equiv\!z\}$. The Pauli matrix of the spin at site
$\alpha\in A$ is given by $\hat{\sigma}_{i}^{(\alpha)}$, and analogously for a
spin $\beta$ from subsystem $B$. Note that the collective operators satisfy
the usual commutation relations $[\hat{S}_{i}^{A},\hat{S}_{j}^{A}%
]=\;$i$\,\hbar\,\varepsilon_{ijk}\,\hat{S}_{k}^{A}$, since $[\hat{\sigma}%
_{i}^{(\alpha)},\hat{\sigma}_{j}^{(\alpha^{\prime})}]=2\,$i$\,\varepsilon
_{ijk}\,\hat{\sigma}_{k}^{(\alpha)}\,\delta_{\alpha\alpha^{\prime}}$.

The spin expectation values and correlations are%
\begin{align}
S_{i}^{A}  &  \equiv\left\langle \!\right.  \hat{S}_{i}^{A}\left.
\!\right\rangle =\frac{\hbar}{2}\,%
%TCIMACRO{\dsum \nolimits_{\alpha\in A}}%
%BeginExpansion
{\displaystyle\sum\nolimits_{\alpha\in A}}
%EndExpansion
\,g_{i}(\alpha)\,,\label{eq S}\\
T_{ij}^{AB}\,  &  \equiv\left\langle \!\right.  \hat{S}_{i}^{A}\hat{S}_{j}%
^{B}\left.  \!\right\rangle =\frac{\hbar^{2}}{4}\,%
%TCIMACRO{\dsum \nolimits_{\alpha\in A,\beta\in B}}%
%BeginExpansion
{\displaystyle\sum\nolimits_{\alpha\in A,\beta\in B}}
%EndExpansion
\,h_{ij}(\alpha,\beta)\,, \label{eq T}%
\end{align}
and analogously for $S_{i}^{B}\equiv\frac{\hbar}{2}\,{\sum\nolimits_{\beta\in
B}}\,g_{i}(\beta)$, where $S_{i}^{A},S_{i}^{B}\!\in\![-\frac{n\hbar}{2}%
,\frac{n\hbar}{2}]$, $T_{ij}^{AB}\!\in\![-\frac{n^{2}\hbar^{2}}{4},\frac
{n^{2}\hbar^{2}}{4}]$. These are only 15 numbers. Here $g_{i}(\alpha)$,
$g_{i}(\beta)$ and $h_{ij}(\alpha,\beta)$ are the (dimensionless) expectation
values and pair correlations of two single spins ($\alpha,\beta$) \textit{to
which it is assumed there is no experimental access}: $g_{i}(\alpha
)\equiv\left\langle \!\right.  \hat{\sigma}_{i}^{(\alpha)}\left.
\!\right\rangle _{\hat{\rho}_{\alpha\beta}}$, $h_{ij}(\alpha,\beta
)\equiv\left\langle \!\right.  \hat{\sigma}_{i}^{(\alpha)}\hat{\sigma}%
_{j}^{(\beta)}\left.  \!\right\rangle _{\hat{\rho}_{\alpha\beta}}$. They are
obtained from the actual $4\!\times\!4$ density matrix%
\begin{align}
\hat{\rho}_{\alpha\beta}  &  \equiv\frac{1}{4}\left[  \rule{0pt}{16pt}%
\!\right.  \mathbf{1}^{(\alpha)}\otimes\mathbf{1}^{(\beta)}+%
%TCIMACRO{\dsum \limits_{k=1}^{3}}%
%BeginExpansion
{\displaystyle\sum\limits_{k=1}^{3}}
%EndExpansion
\,g_{k}(\alpha)\,\hat{\sigma}_{k}^{(\alpha)}\otimes\mathbf{1}^{(\beta
)}\label{rhoalphabeta}\\
&  +%
%TCIMACRO{\dsum \limits_{l=1}^{3}}%
%BeginExpansion
{\displaystyle\sum\limits_{l=1}^{3}}
%EndExpansion
\,\mathbf{1}^{(\alpha)}\otimes g_{l}(\beta)\,\hat{\sigma}_{l}^{(\beta)}+%
%TCIMACRO{\dsum \limits_{k,l=1}^{3}}%
%BeginExpansion
{\displaystyle\sum\limits_{k,l=1}^{3}}
%EndExpansion
h_{kl}(\alpha,\beta)\,\hat{\sigma}_{k}^{(\alpha)}\otimes\hat{\sigma}%
_{l}^{(\beta)}\left.  \rule{0pt}{16pt}\!\right]  \!,\nonumber
\end{align}
where $\mathbf{1}^{(\alpha)}$ is the $2\!\times\!2$ identity matrix in the
Hilbert space of spin $\alpha$.

Out of the experimentally accessible quantities (\ref{eq S}) and (\ref{eq T})
we will construct a $4\!\times\!4$ density matrix of \textit{two virtual
qubits} which describes the collective properties of the two spin sets. Its
\textit{a priori} justification is (i) that a general treatment of the problem
between two large samples of spins is intractable because of the high
dimensionality, and (ii) that we have a fully developed theory of entanglement
for two-qubit systems. Therefore, this approach is a natural (and successful)
way to say something at all about the entanglement between two spin systems if
only collective observables are measured.

We first introduce the normalized (dimensionless) average subsystem
expectation values (magnetization per particle) and correlations:
\begin{align}
s_{i}^{a}  &  \equiv\frac{1}{n}\,%
%TCIMACRO{\dsum \nolimits_{\alpha\in A}}%
%BeginExpansion
{\displaystyle\sum\nolimits_{\alpha\in A}}
%EndExpansion
\,g_{i}(\alpha)=\frac{2}{n\hbar}\,S_{i}^{A}\,,\label{si}\\
t_{ij}^{ab}  &  \equiv\frac{1}{n^{2}}\,%
%TCIMACRO{\dsum \nolimits_{\alpha\in A,\beta\in B}}%
%BeginExpansion
{\displaystyle\sum\nolimits_{\alpha\in A,\beta\in B}}
%EndExpansion
\,h_{ij}(\alpha,\beta)=\frac{4}{n^{2}\hbar^{2}}\,T_{ij}^{AB}\,, \label{tij}%
\end{align}
where $s_{i}^{a},t_{ij}^{ab}\in\lbrack-1,1]$. These are the coefficients of
the virtual density matrix:
\begin{align}
\hat{\rho}_{ab}  &  \equiv\frac{1}{4}\left[  \rule{0pt}{16pt}\!\right.
\mathbf{1}^{a}\otimes\mathbf{1}^{b}+%
%TCIMACRO{\dsum \limits_{k=1}^{3}}%
%BeginExpansion
{\displaystyle\sum\limits_{k=1}^{3}}
%EndExpansion
\,s_{k}^{a}\,\hat{\sigma}_{k}^{a}\otimes\mathbf{1}^{b}+\label{rho12}\\
&  \quad\,+%
%TCIMACRO{\dsum \limits_{l=1}^{3}}%
%BeginExpansion
{\displaystyle\sum\limits_{l=1}^{3}}
%EndExpansion
\,\mathbf{1}^{a}\otimes s_{l}^{b}\,\hat{\sigma}_{l}^{b}+%
%TCIMACRO{\dsum \limits_{k,l=1}^{3}}%
%BeginExpansion
{\displaystyle\sum\limits_{k,l=1}^{3}}
%EndExpansion
t_{kl}^{ab}\,\hat{\sigma}_{k}^{a}\otimes\hat{\sigma}_{l}^{b}\left.
\rule{0pt}{16pt}\!\right]  \!,\nonumber
\end{align}
with $a$ denoting the first and $b$ the second virtual collective qubit,
associated with subsystems $A$ and $B$, respectively. Here, $\mathbf{1}^{a}$,
$\mathbf{1}^{b}$, $\hat{\sigma}_{k}^{a}$ and $\hat{\sigma}_{l}^{b}$ are
$2\!\times\!2$ identity and Pauli matrices for the collective qubits $a$ and
$b$.

The question is whether the density matrix (\ref{rho12}) is positive
semi-definite, i.e., whether it is a physical state of two qubits. The answer
is affirmative and the proof follows from consideration of an equal-weight
statistical mixture of \textit{one} qubit pair which can be in any of the
$n^{2}$ states $\hat{\rho}_{\alpha\beta}$. The density matrix of this mixture
is the mixture of density matrices of all possible pairs ($\alpha,\beta$):
$\hat{\rho}_{\text{mix}}=\frac{1}{n^{2}}\,{\sum\nolimits_{\alpha,\beta}}%
\,\hat{\rho}_{\alpha\beta}$. It can easily be seen that $\hat{\rho
}_{\text{mix}}$ is equal to $\hat{\rho}_{ab}$ as both are uniquely determined
by the same expectations and correlations: $\langle\hat{\sigma}_{i}^{(\alpha
)}\rangle_{\hat{\rho}_{\text{mix}}}=\langle\hat{\sigma}_{i}^{a}\rangle
_{\hat{\rho}_{ab}}=s_{i}^{a}=\frac{2}{n\hbar}\,S_{i}^{A}$ and $\langle
\hat{\sigma}_{i}^{(\alpha)}\hat{\sigma}_{j}^{(\beta)}\rangle_{\hat{\rho
}_{\text{mix}}}=\langle\hat{\sigma}_{i}^{a}\hat{\sigma}_{j}^{b}\rangle
_{\hat{\rho}_{ab}}=t_{ij}^{ab}=\frac{4}{n^{2}\hbar^{2}}\,T_{ij}^{AB}$. Thus,
$\hat{\rho}_{ab}$ is a density matrix. Note that without the normalizations as
given in (\ref{si}) and (\ref{tij}), the method would not work.

Encapsulated in the following two propositions, we relate the entanglement
properties of the virtual qubits to those of the spin samples.

\textit{Proposition 1}. \textit{For any entanglement measure }$E$\textit{ that
is convex on the set of density matrices the entanglement of the virtual
density matrix }$E_{ab}\equiv E(\hat{\rho}_{ab})$\textit{ is a lower bound for
the average entanglement between all pairs }$\bar{E}_{\alpha\beta}\equiv
\tfrac{1}{n^{2}}\,\sum\nolimits_{\alpha,\beta}\,E(\hat{\rho}_{\alpha\beta}%
)$\textit{.} This is an immediate consequence of the convexity of $E$:
$E_{ab}=E(\tfrac{1}{n^{2}}\,{\sum\nolimits_{\alpha,\beta}}\,\hat{\rho}%
_{\alpha\beta})\leq\tfrac{1}{n^{2}}\,{\sum\nolimits_{\alpha,\beta}}%
\,E(\hat{\rho}_{\alpha\beta})$.

Remarks: First, the result holds for entanglement measures that are convex. In
certain cases this is directly implied by the definition of the entanglement
measure for mixed states, which involves a convex roof $E(\hat{\rho}%
)\equiv\text{min}_{p_{i},\psi_{i}}\sum_{i}p_{i}\,E(|\psi_{i}\rangle\langle
\psi_{i}|)$, where the minimization is taken over those probabilities
$\{p_{i}\}$ and pure states $|\psi_{i}\rangle$ that realize the density matrix
$\hat{\rho}=\sum_{i}p_{i}|\psi_{i}\rangle\langle\psi_{i}|$ and $E(|\psi
_{i}\rangle\langle\psi_{i}|)$ is the entanglement measure of the pure state
$|\psi_{i}\rangle$. Second, the proposition implies that if $E_{ab}\!>\!0$
then at least for one pair we must have $E(\hat{\rho}_{\alpha\beta})\!>\!0$.
Thus, a non-zero value of $E_{ab}$ is a sufficient condition for entanglement
between the two samples $A$ and $B$. Third, the maximal pairwise
concurrence~\cite{Wootters} for symmetric states is found to be $2/n$ and is
achieved for the W-state~\cite{Koachi}. It is conjectured that this remains
valid also when the symmetry constraint is removed. This suggests $E_{ab}%
\leq\bar{E}_{\alpha\beta}\leq2/n$, if concurrence is used as an entanglement
measure. The existence of this upper bound can be seen as a consequence of the
monogamy of entanglement.

We refer to $E_{ab}$ as \textit{pairwise collective entanglement} as it is
determined solely by the expectation and correlation values of the collective
spin observables. The question arises: Under what conditions is $E_{ab}$ equal
to the average entanglement $\bar{E}_{\alpha\beta}$? Identifying systems for
which the equality holds, would allow feasible experimental determination of
the entanglement distribution in large samples by observation of their
macroscopic properties only. It can easily be seen that, if the state is
symmetric under exchange of particles \textit{within} each of the samples, one
has $E_{ab}=\bar{E}_{\alpha\beta}=E(\hat{\rho}_{\alpha\beta})$ for every pair
of particles $(\alpha,\beta)$. In what follows, we identify an important class
of systems for which $E_{ab}\!=\!\bar{E}_{\alpha\beta}$ though the
corresponding states need \textit{not} be symmetric under exchange of particles.

\textit{Proposition 2}. Consider a system (i) with $g_{z}(\alpha)=g_{z}^{A}$
for all $\alpha\in A$ and $g_{z}(\beta)=g_{z}^{B}$ for all $\beta\in B$
(translational invariance within the subsystems), (ii) with $h_{xx}%
(\alpha,\beta)=\varepsilon\,h_{yy}(\alpha,\beta)$ with $\varepsilon
=\;$const$\;=+1$ or $-1$ (all pairs are in absolute value equally correlated
in the $x$ and $y$-direction, where this correlation may be different in size
for different pairs), (iii) with constant sign of the $z$-correlations, i.e.,
sgn$[h_{zz}(\alpha,\beta)]=-\varepsilon$ for all $(\alpha,\beta)$, and (iv)
where all the remaining expectation values and correlations ($g_{x}$, $g_{y}$,
$h_{ij}$ with $i\neq j$) are zero. \textit{Non-vanishing average entanglement
}$\bar{E}_{\alpha\beta}$\textit{ as measured by the negativity~}%
\cite{Zycz1998}\textit{ is equal to the pairwise collective entanglement
}$E_{ab}$\textit{, if and only if the correlation functions }$h_{xx}%
(\alpha,\beta)$ and $h_{yy}(\alpha,\beta)$\textit{ are each constant for all
pairs and all pairs have a non-positive eigenvalue of their partial transposed
density matrix.}

The negativity~\cite{Zycz1998} of a density matrix $\hat{\rho}$ is defined as
$E(\hat{\rho})\equiv(\text{Tr}|\hat{\rho}^{\text{pT}}|-1)/2$, where
$\text{Tr}|\hat{\rho}^{\text{pT}}|$ stands for the trace norm of the partially
transposed density matrix $\hat{\rho}^{\text{pT}}$. Hence the negativity is
equal to the modulus of the sum of the negative eigenvalues of $\rho
^{\text{pT}}$.

It is important to stress that proposition 2 holds also for states that do not
need to be totally symmetric, i.e., the $h_{zz}(\alpha,\beta)$ may be
different for different pairs of particles. In general, under the above
symmetry, the state of the virtual qubit pair is of the form%
\begin{equation}
\hat{\rho}_{ab}={\small
\begin{pmatrix}
u_{+++} & 0 & 0 & v_{-}\\
0 & u_{+--} & v_{+} & 0\\
0 & v_{+} & u_{-+-} & 0\\
v_{-} & 0 & 0 & u_{--+}%
\end{pmatrix}
\!,}%
\end{equation}
where $u_{\pm\pm\pm}\equiv\frac{1}{4}\,(1\pm s_{z}^{A}\pm s_{z}^{B}\pm
t_{zz}^{ab})$, $v_{\pm}\equiv\frac{1}{4}\,t_{xx}^{ab}\,(1\pm\varepsilon)$. The
state of an arbitrary pair $(\alpha,\beta)$ of particles has a similar
structure. For example, the state of an arbitrary pair of particles extracted
from a spin chain with $xxz$ Heisenberg interaction has such a form. The proof
of proposition 2 is given in the appendix.

We illustrate the method with some explicit examples.

\textit{1. Dicke states}. We consider the Dicke state (generalized W-state)%
\begin{equation}
\left|  N;k\right\rangle \equiv\left(
%TCIMACRO{\QTATOP{N}{k}}%
%BeginExpansion
\genfrac{}{}{0pt}{1}{N}{k}%
%EndExpansion
\right)  ^{\!-1/2}\,\hat{P}_{S}\,|\underbrace{0...0}_{N-k}\underbrace
{1...1}_{k}\rangle\label{W state}%
\end{equation}
with $N\geq2$ spins, $k$ excitations $|1\rangle$ and $N-k$ non-excited spins
$|0\rangle$, where $0\leq k\leq N$. $\hat{P}_{S}$ is the symmetrization
operator. Within the system we consider two subsystems $A$ and $B$ each of
size $n$. Because of the total symmetry of the state one has $E_{ab}=\bar
{E}_{\alpha\beta}=E(\hat{\rho}_{\alpha\beta})$ for any size $n$. Only for the
cases where just a single spin ($k=0$) or all spins are excited ($k=N$), there
is no entanglement between two arbitrary pairs or arbitrary sized blocks,
respectively~\cite{Vedr2004}. The global maximum of the entanglement $E_{ab}$,
measured by the negativity, is reached for $k=\frac{N}{2}$ and its value is
$E_{ab}^{\text{max}}=\frac{1}{2}\,\frac{1}{N-1}$, for all $n$. It vanishes
only in the limit $N\rightarrow\infty$.

\textit{2.\ Generalized singlet states}. The two subsystems $A$ and $B$, each
forming a spin $s=\frac{n}{2}$, are in a generalized singlet state:%
\begin{equation}
\left|  \psi\right\rangle =\frac{1}{\sqrt{2\,s+1}}\,%
%TCIMACRO{\dsum \limits_{m=-s}^{s}}%
%BeginExpansion
{\displaystyle\sum\limits_{m=-s}^{s}}
%EndExpansion
(-1)^{s-m}\left|  m\right\rangle _{A}\left|  -m\right\rangle _{B},
\label{general singlet}%
\end{equation}
where $\left|  m\right\rangle =\left|  2\,s;s+m\right\rangle $\ denotes the
eigenstates of the spin operator's $z$-component. The collective two-qubit
coefficients are $t_{ii}^{ab}=-\frac{n+2}{3n}$ and the $s_{i}^{a,b}$ and
$t_{ij}^{ab}$ with $i\neq j$ are all zero. The collective entanglement
(negativity) is $E_{ab}=\frac{1}{2n}=\frac{1}{4s}$ which is non-zero for all
sizes $n$ of the subsystems and vanishes only in the limit $n\rightarrow
\infty$.

\begin{figure}[t]
\begin{center}
\includegraphics{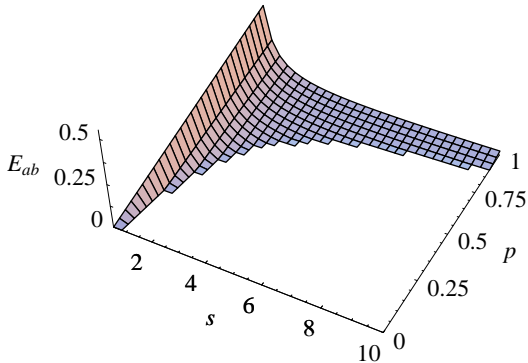}
\end{center}
\par
\vspace{-0.75cm}\caption{(Color online.) Entanglement $E_{ab}$ between two
collective spins in the generalized singlet state with an admixture of
non-symmetric noise (\ref{singlet plus z-corr}) as a function of spin length
$s$ and proportion $p$ of the singlet state in the mixture. The entanglement
is non-zero for sufficiently large $p$ and decreases inversely proportionally
to $s$.}%
\label{fig plots}%
\end{figure}

\textit{3. Generalized singlet state with an admixture of non-symmetric
correlations}. Consider the state%
\begin{equation}
p\left|  \psi\right\rangle \!\left\langle \psi\right|  +(1-p)\,%
%TCIMACRO{\dbigotimes \limits_{\alpha=1}^{n}}%
%BeginExpansion
{\displaystyle\bigotimes\limits_{\alpha=1}^{n}}
%EndExpansion
\,\dfrac{1}{2}\left(  \left|  01\right\rangle _{\alpha,\beta=\alpha
}\!\left\langle 10\right|  +\left|  10\right\rangle _{\alpha,\beta=\alpha
}\!\left\langle 01\right|  \right)  \label{singlet plus z-corr}%
\end{equation}
with $p\in\lbrack0,1]$. This is a mixture of the generalized singlet state
(\ref{general singlet}) and $n$ perfectly $z$-correlated pairs ($\alpha
,\beta\!=\!\alpha$). This state is \textit{not} symmetric under particle
exchange in the $zz$-correlations. The expectation values $s_{i}^{a,b}$ and
correlations $t_{ij}^{ab}$ with $i\neq j$ remain zero. The correlations
$t_{xx}=t_{yy}$ are reduced by a factor $p$ compared to those of the state
(\ref{general singlet}). The correlations in $z$-direction, however, are
modified and read $t_{zz}^{ab}=-p\,\frac{n-1}{3n}-\frac{1}{n}$. Therefore,
there is a critical number of particles $n_{c}\equiv\left\lceil \!\right.
\frac{1+p}{1-p}\left.  \!\right\rceil $ beyond which there is no collective
pairwise entanglement. Only for $n<n_{c}$ we have $E_{ab}=\frac{1}{4}%
\frac{1+p-n\,(1-p)}{n}>0$. Note that (\ref{singlet plus z-corr}) is in
accordance with proposition 2 (appendix) and thus $E_{ab}=\bar{E}_{\alpha
\beta}$. Figure~\ref{fig plots}\ shows $E_{ab}$ as a function of the spin
length $s=\frac{n}{2}$ and the mixing parameter $p$. $E_{ab}$ is non-zero in
regions where $p>\frac{2s-1}{2s+1}$ and decreases inversely proportionally to
$s$.

Our method can be generalized straightforwardly to define multi-partite
entanglement of $M$ collective spins belonging to $M$ separated samples
$A_{1},...,A_{M}$, each containing a large number of spins $n$. Any convex
multi-partite entanglement measure (e.g., $M$-way tangle) which is applied to
the corresponding collective matrix of the $M$ virtual qubits gives a
\textit{lower bound for the average multi-partite entanglement}, obeying the
usual constraints for entanglement sharing such as the
Coffman--Kundu--Wootters inequality~\cite{Coffman}.

Importantly, in the examples considered above our entanglement measure scales
at most with $1/n$ and vanishes in the limit of infinitely large subsystem
sizes $n$. This is a generic property that follows from the commutation
relation for \textit{normalized} spins in this limit. Taking $\hat{s}%
_{i}\equiv\frac{1}{n}\,\hat{S}_{i}=\frac{\hbar}{2n}\sum_{\alpha}\hat{\sigma
}_{i}^{(\alpha)}$ one obtains $\lim_{n\rightarrow\infty}[\hat{s}_{x},\hat
{s}_{y}]=\lim_{n\rightarrow\infty}\,$i$\,\frac{\hbar}{2n}\,\hat{s}_{z}=0$.
This is sometimes interpreted as suggesting that averaged collective
observables, like the magnetization per particle, represent ''macroscopic'' or
classical-like, properties of samples. Note, however, that for any $n$ there
are $n^{2}$ pairs between the subsystems so that the number of pairs
multiplied by the pairwise collective entanglement can scale with $n$, showing
the existence of entanglement for arbitrarily large $n$.

\textit{Conclusion}. In a recent work it was shown that macroscopic properties
such as magnetic susceptibility can reveal entanglement \textit{within}
macroscopic samples~\cite{Wiesniak}. The present work can be viewed as in a
way complementary as it demonstrates that macroscopic properties (collective
spin properties and their correlations) can reveal the entanglement
distribution \textit{between} two or more macroscopic samples. On the
fundamental side, our method demonstrates that there is no reason in principle
why purely quantum correlations could not have an effect on the global
properties of objects. On the practical side, it enables us to characterize
the structure of entanglement in large spin systems by performing only a few
feasible measurements of their collective properties, independent of the
symmetry and mixedness of the state.

This work was supported by the Austrian Science Foundation, Proj.\ SFB
(No.~1506), the Europ.\ Commission, Proj.\ QAP (No.~015846), and the British
Council in Austria.

\textit{Appendix. Proof of proposition 2}. Depending on the sign $\varepsilon$
of the $zz$-correlations, only one eigenvalue of $\hat{\rho}_{\alpha\beta
}^{\text{pT}}$ and $\hat{\rho}_{ab}^{\text{pT}}$, respectively, can be
negative:%
\begin{align}
\mu_{\alpha\beta}  &  =\dfrac{1}{4}\!\left[  1-\!\sqrt{(g_{z}^{A}%
+\varepsilon\,g_{z}^{B})^{2}+4\,h_{xx}^{2}(\alpha,\beta)}+\varepsilon
\,h_{zz}(\alpha,\beta)\right]  \!\!,\nonumber\\
\nu_{ab}  &  =\dfrac{1}{4}\!\left[  1-\!\sqrt{(s_{z}^{a}+\varepsilon
\,s_{z}^{b})^{2}+4\,(t_{xx}^{ab})^{2}}+\varepsilon\,t_{zz}^{ab}\right]  \!\!.
\label{evn1}%
\end{align}
The corresponding negativities are given by $E_{\alpha\beta}=|\!\min
(0,\mu_{\alpha\beta})|$ and $E_{ab}=|\!\min(0,\nu_{ab})|$. One can express
$\nu_{ab}$ as given by%
\begin{equation}
\nu_{ab}=\bar{\mu}+\Delta\,,
\end{equation}
where $\bar{\mu}\equiv\frac{1}{n^{2}}\,{\sum\nolimits_{\alpha,\beta}}%
\,\mu_{\alpha\beta}$ and $\Delta\equiv\frac{1}{4n^{2}}\,{\sum\nolimits_{\alpha
,\beta}}[(g_{z}^{A}+\varepsilon\,g_{z}^{B})^{2}+4\,h_{xx}^{2}(\alpha
,\beta)]^{1/2}-\frac{1}{4}[(s_{z}^{a}+\varepsilon\,s_{z}^{b})^{2}%
+4\,(t_{xx}^{ab})^{2}]^{1/2}$. The quantity $\Delta$ is the difference between
the entanglement measures $\bar{E}_{\alpha\beta}$ and $E_{ab}$, i.e.,
$E_{ab}=\bar{E}_{\alpha\beta}-\Delta$, for the case that $\nu_{ab}\leq0$ and
$\bar{E}_{\alpha\beta}\equiv\tfrac{1}{n^{2}}\,{\sum\nolimits_{\alpha,\beta}%
}\,|\!\min(0,\mu_{\alpha\beta})|=|\!\min(0,\bar{\mu})|$. This is true, if and
only if $\mu_{\alpha\beta}\leq0$ for all ($\alpha,\beta$), i.e., all pairs are
either entangled or have eigenvalue zero. According to proposition 1, $\Delta$
is non-negative, i.e.,%
\begin{equation}
\sqrt{c^{2}+4\,(t_{xx}^{ab})^{2}}\leq\tfrac{1}{n^{2}}\,%
%TCIMACRO{\tsum \nolimits_{\alpha,\beta}}%
%BeginExpansion
{\textstyle\sum\nolimits_{\alpha,\beta}}
%EndExpansion
\,\sqrt{c^{2}+4\,h_{xx}^{2}(\alpha,\beta)}\,. \label{red1}%
\end{equation}
Here we abbreviated $c\equiv g_{z}^{A}+\varepsilon\,g_{z}^{B}=s_{z}%
^{a}+\varepsilon\,s_{z}^{b}$, where the latter equal sign is due to
(\ref{si}). Inequality (\ref{red1}) becomes an equality, i.e., $\Delta=0$, if
and only if $h_{xx}(\alpha,\beta)$ is the same for all pairs $(\alpha,\beta)$
such that $t_{xx}^{ab}=h_{xx}$. Therefore, the pairwise collective
entanglement $E_{ab}$ equals the average entanglement $\bar{E}_{\alpha\beta}$,
if and only if for all individual pairs $\mu_{\alpha\beta}\leq0$ and
$h_{xx}(\alpha,\beta)=\varepsilon\,h_{yy}(\alpha,\beta)=\;$const for all
pairs. $\square$

\end{document}